\documentclass{ws-procs9x6}

\begin{document}

\newcommand{\no}{\nonumber}
\newcommand{\nn}{\nonumber\\}
\newcommand{\N}{{\bf N}}
\newcommand{\Z}{{\bf Z}}
\newcommand{\Q}{{\bf Q}}
\newcommand{\R}{{\bf R}}
\newcommand{\C}{{\bf C}}
\newcommand{\cA}{{\cal A}}
\newcommand{\cB}{{\cal B}}
\newcommand{\cD}{{\cal D}}
\newcommand{\cE}{{\cal E}}
\newcommand{\cF}{{\cal F}}
\newcommand{\cG}{{\cal G}}
\newcommand{\cH}{{\cal H}}
\newcommand{\cI}{{\cal I}}
\newcommand{\cJ}{{\cal J}}
\newcommand{\cK}{{\cal K}}
\newcommand{\cL}{{\cal L}}
\newcommand{\cM}{{\cal M}}
\newcommand{\cN}{{\cal N}}
\newcommand{\cO}{{\cal O}}
\newcommand{\cP}{{\cal P}}
\newcommand{\cQ}{{\cal Q}}
\newcommand{\cR}{{\cal R}}
\newcommand{\cS}{{\cal S}}
\newcommand{\cT}{{\cal T}}
\newcommand{\cU}{{\cal U}}
\newcommand{\cV}{{\cal V}}
\newcommand{\cW}{{\cal W}}
\newcommand{\cX}{{\cal X}}
\newcommand{\cY}{{\cal Y}}
\newcommand{\cZ}{{\cal Z}}
\newcommand{\na}{\nabla}
\newcommand{\ve}{\varepsilon}
\newcommand{\vphi}{\varphi}
\newcommand{\bphi}{\bar{\phi}}
\newcommand{\bpsi}{\bar{\psi}}
\newcommand{\hpsi}{\hat{\psi}}
\newcommand{\del}{\partial}
\newcommand{\vvert}{\Big{\vert}}
\newcommand{\bra}{\langle}
\newcommand{\ket}{\rangle}
\newcommand{\vev}[1]{\langle #1\rangle}
\newcommand{\Bra}{\Big{\langle}}
\newcommand{\Ket}{\Big{\rangle}}
\newcommand{\Vev}[1]{\Big{\langle}#1\Big{rangle}}
\newcommand{\gr}{\mbox{grad}\,}
\newcommand{\di}{\mbox{div}\,}
\newcommand{\rot}{\mbox{rot}\,}
\newcommand{\diag}{\mbox{diag}\,}
\newcommand{\rk}{{\rm rank}\,}
\newcommand{\tr}{{\rm tr}\,}
\newcommand{\Tr}{{\rm Tr}\,}
\newcommand{\trt}{{\rm tr}_2}
\newcommand{\Trk}{{\rm Tr}_k}
\newcommand{\TrN}{{\rm Tr}_N}
\newcommand{\Hom}{{\rm Hom}\,}
\newcommand{\Ker}{{\rm Ker}\,}
\newcommand{\re}{{\rm Re}\,}
\newcommand{\im}{{\rm Im}\,}
\newcommand{\Det}{{\rm Det}\,}
\newcommand{\all}{{~}^{\forall}}
\newcommand{\ex}{{~}^{\exist}}
\newcommand{\ch}{\check}
\newcommand{\da}{\dagger}
\newcommand{\pr}{\prime}
\newcommand{\til}{\tilde}
\newcommand{\wti}{\widetilde}
\newcommand{\wha}{\widehat}
\newcommand{\rar}{\rightarrow}
\newcommand{\lar}{\leftarrow}
\newcommand{\lra}{\leftrightarrow}
\newcommand{\uar}{\uparrow}
\newcommand{\dar}{\downarrow}
\newcommand{\Rar}{\Rightarrow}
\newcommand{\Lar}{\Leftarrow}
\newcommand{\Lra}{\Leftrightarrow}
\newcommand{\uda}{\updownarrow}
\newcommand{\longr}{\longrightarrow}
\newcommand{\longl}{\longleftarrow}
\newcommand{\longlr}{\longleftrightarrow}
\newcommand{\Longr}{\Longrightarrow}
\newcommand{\Longl}{\Longleftarrow}
\newcommand{\Longlr}{\Longleftrightarrow}
\newcommand{\Uda}{\Updownarrow}
\newcommand{\nsa}{{\ooalign{\hfil$\nearrow$\hfil\crcr$\swarrow$}}}
\newcommand{\sna}{{\ooalign{\hfil$\searrow$\hfil\crcr$\nwarrow$}}}
\newcommand{\map}{\mapsto}
\newcommand{\lmap}{\longmapsto}
\newcommand{\ti}{\times}
\newcommand{\op}{\oplus}
\newcommand{\ot}{\otimes}
\newcommand{\we}{\wedge}
\newcommand{\ap}{\approx}
\newcommand{\st}{\stackrel}
\newcommand{\unb}{\underbrace}
\newcommand{\unl}{\underline}
\newcommand{\lab}{\label}
\newcommand{\fr}{\frac}
\newcommand{\half}{\frac{1}{2}}
\newcommand{\qua}{\frac{1}{4}}
\newcommand{\scr}{\scriptsize}
\newcommand{\Dsl}{\mbox{\ooalign{\hfil/\hfil\crcr$D$}}}
\newcommand{\dbar}{{\ooalign{\hfil$^{\,-}$\hfil\crcr$d$}}}
\newcommand{\db}[1]{{\ooalign{\hfil$^{\,-#1}$\hfil\crcr$d$}}}
\newcommand{\intb}{{\ooalign{\hfil$-$\hfil\crcr$\dis\int$}}}
\newcommand{\ointb}{{\ooalign{\hfil$-$\hfil\crcr$\dis\oint$}}}
\newcommand{\wwr}{\mbox{\ooalign{\hfil$\,\wr$\hfil\crcr$\wr$}}}
\newcommand{\dis}{\displaystyle}
\newcommand{\mn}{{\mu\nu}}
\newcommand{\eb}{\bar{e}}
\newcommand{\ib}{\bar{\imath}}
\newcommand{\jb}{\bar{\jmath}}
\newcommand{\kb}{\bar{k}}
\newcommand{\lb}{\bar{l}}
\newcommand{\qb}{\bar{q}}
\newcommand{\ub}{\bar{u}}
\newcommand{\wb}{\bar{w}}
\newcommand{\zb}{\bar{z}}
\newcommand{\cDb}{\bar{\cD}}
\newcommand{\nab}{\bar{\na}}
\newcommand{\ah}{\hat{a}}
\newcommand{\fh}{\hat{f}}
\newcommand{\gh}{\hat{g}}
\newcommand{\mh}{{\hat{\mu}}}
\newcommand{\nh}{{\hat{\nu}}}
\newcommand{\sh}{\hat{s}}
\newcommand{\uh}{\hat{u}}
\newcommand{\vh}{\hat{v}}
\newcommand{\xh}{\hat{x}}
\newcommand{\zh}{\hat{z}}
\newcommand{\Ah}{\hat{A}}
\newcommand{\Bh}{\hat{B}}
\newcommand{\Dh}{\hat{D}}
\newcommand{\Fh}{\hat{F}}
\newcommand{\Ph}{\hat{P}}
\newcommand{\Sh}{\hat{S}}
\newcommand{\Th}{\hat{T}}
\newcommand{\Uh}{\hat{U}}
\newcommand{\Vh}{\hat{V}}
\newcommand{\phih}{\hat{\phi}}
\newcommand{\Phih}{\hat{\Phi}}
\newcommand{\zbh}{\hat{\bar{z}}}
\newcommand{\xih}{\hat{\xi}}
\newcommand{\nah}{\hat{\nabla}}
\newcommand{\delh}{\hat{\partial}}
\newcommand{\whn}{\widehat{\nabla}}
\newcommand{\whA}{\widehat{A}}
\newcommand{\whd}{\widehat{D}}
\newcommand{\wht}{\widehat{T}}
\newcommand{\whf}{\widehat{\cF}}
\newcommand{\whg}{\widehat{G}}
\newcommand{\wtph}{\widetilde{\phi}}
\newcommand{\wtps}{\widetilde{\psi}}
\newcommand{\wtc}{\widetilde{\chi}}
\newcommand{\wtx}{\widetilde{\xi}}
\newcommand{\vs}{\vspace*{0.3cm}\noindent}


\title{Noncommutative Solitons\\ and Integrable Systems
\footnote{\uppercase{T}his work was supported in part by
\uppercase{JSPS} \uppercase{R}esearch
\uppercase{F}ellowships for \uppercase{Y}oung
\uppercase{S}cientists (\#0310363)
and the
\uppercase{D}aiko
\uppercase{F}oundation (\#9095).
}}

\author{Masashi HAMANAKA}

\address{Graduate School of Mathematics, Nagoya University,\\
Chikusa-ku, Nagoya, 464-8602, JAPAN \footnote{\uppercase{T}he
author visits the \uppercase{K}orea \uppercase{I}nstitute for
\uppercase{A}dvanced \uppercase{S}tudy (\uppercase{KIAS})
 from 27 \uppercase{M}arch to 2 \uppercase{A}pril, 2005
(\uppercase{P}reprint number: \uppercase{KIAS-P}05026),
and \uppercase{M}athematical \uppercase{I}nstitute, 
\uppercase{U}niversity of \uppercase{O}xford
from 16 \uppercase{A}ugust, 2005 to 15 \uppercase{A}ugust, 2006.}\\
E-mail: hamanaka@math.nagoya-u.ac.jp}

\maketitle

\abstracts{ We review recent developments of soliton theories and
integrable systems on noncommutative spaces. The former part is a
brief review of noncommutative gauge theories focusing on ADHM
construction of noncommutative instantons. The latter part is a
report on recent results of existence of infinite conserved
densities and exact multi-soliton solutions for noncommutative
Gelfand-Dickey hierarchies. Some examples of noncommutative Ward's
conjecture are also presented. Finally, we discuss future
directions on noncommutative Sato's theories and twistor theories.
}

\section{Introduction}

Non-Commutative (NC) extension of field theories has been studied
intensively for the last several years. NC gauge theories are
equivalent to ordinary gauge theories in the presence of
background magnetic fields and succeeded in revealing various
aspects of them. (For reviews, see e.g.
\cite{Chu,DoNe,K.Ito,KoSc,Szabo}.) NC solitons especially play
important roles in the study of D-brane dynamics, such as the
confirmation of Sen's conjecture on tachyon condensation. (For
reviews, see e.g. \cite{Harvey2,Sen}.) One of the distinguished
features of NC theories is resolution of singularities. This gives
rise to various new physical objects such as U(1) instantons and
makes it possible to analyze singular configurations as usual.
(For a review, see my Ph.D thesis \cite{Hamanaka3}.)

NC extension of integrable equations such as the KdV equation is
also one of the hot topics. These equations imply no gauge field
and NC extension of them perhaps might have no physical picture or
no good property on integrabilities. To make matters worse, NC
extension of $(1+1)$-dimensional equations introduces infinite
number of time derivatives, which makes it hard to discuss or
define the integrability. However, some of them actually possess
integrable-like properties, such as the existence of infinite number of
conserved quantities and the linearizability which are widely
accepted as definition of complete integrability of equations.
Furthermore, a few of them can be derived from NC (anti-)self-dual
Yang-Mills (ASDYM) equations by suitable reductions. This fact may
give some physical meanings and good properties to the
lower-dimensional NC field equations and makes us expect that
Ward's conjecture \cite{Ward} still holds on NC spaces. So far,
however, those equations have been examined one by one. Now it is
time to discuss the geometrical and physical origin of the special
properties and integrabilities, in more general framework.

We would like to propose the following study programs as future
directions:
\begin{itemize}
\item Construction of NC twistor theory
\item Confirmation of NC Ward's conjecture
\item Completion of NC Sato's theory
\end{itemize}

Twistor theory \cite{Penrose} is the most essential framework in
the study of integrability of ASD Yang-Mills(-Higgs) equations.
(See, e.g. \cite{MaWo,WaWe}.) NC extension of twistor theories are
already discussed by several authors, e.g.
\cite{Hannabuss,IhUh,KKO,LePo3,Takasaki}. This would lays the
geometrical foundation of integrabilities of ASD YM(H) equations.

NC Ward's conjecture is very important to give physical pictures to
lower-dimensional integrable equations and to make it possible to
apply analysis of NC solitons to that of the corresponding
D-branes. Origin of the integrable-like properties would be also
revealed from the viewpoints of NC twistor theory and preserved
supersymmetry in the D-brane systems.

Sato's theory is known to be one of the most beautiful theories of
solitons and reveals essential aspects of the integrability, such
as, the construction of exact multi-soliton solutions, the
structure of the solution space, the existence of infinite
conserved quantities, and the hidden symmetry of them, on
commutative spaces. So it is reasonable to extend Sato's theory
onto NC spaces in order to clarify various
integrable-like aspects directly.

In this article, we report recent developments of NC extension of
soliton theories and integrable systems focusing on NC ADHM
construction and NC Sato's theory. As recent results
\cite{Hamanaka4,Hamanaka5}, we prove the existence of infinite
conserved quantities and exact multi-soliton solutions for
Gelfand-Dickey hierarchies on NC spaces and give the explicit
representations with both space-space and space-time
noncommutativities. Our results include NC versions of KP, KdV,
Boussinesq, coupled KdV, Sawada-Kotera, modified KdV equations and
so on.

\section{NC Instantons and BPS Monopoles}

\subsection{NC Gauge Theories}

NC spaces are defined
by the noncommutativity of the coordinates:
\begin{eqnarray}
\label{nc_coord}
[x^i,x^j]=i\theta^{ij},
\end{eqnarray}
where $\theta^{ij}$ are real constants and
called the {\it NC parameters}.
This relation looks like the canonical commutation
relation in quantum mechanics
and leads to ``space-space uncertainty relation.''
Hence the singularity which exists on commutative spaces
could resolve on NC spaces.
This is one of the prominent features of NC
field theories and yields various new physical objects.

NC field theories are given by the exchange of ordinary products
in the commutative field theories for the star-products and
realized as deformed theories from the commutative ones. In this
context, they are often called the {\it NC-deformed theories}.

The star-product is defined for ordinary fields on commutative
spaces. For Euclidean spaces, it is explicitly given by
\begin{eqnarray}
f\star g(x)&:=&\mbox{exp }
\left(\frac{i}{2}\theta^{ij} \partial^{(x^{\prime})}_i
\partial^{(x^{\prime\prime})}_j \right)
f(x^\prime)g(x^{\prime\prime})\Big{\vert}_{x^{\prime}
=x^{\prime\prime}=x}\nonumber\\
&=&f(x)g(x)+\frac{i}{2}\theta^{ij}\partial_i f(x)\partial_j g(x)
+O (\theta^2),
\label{star}
\end{eqnarray}
where $\del_i^{(x^\prime)}:=\del/\del x^{\prime i}$
and so on.
This explicit representation is known
as the {\it Moyal product} \cite{Groenewold,Moyal}.

The star-product has associativity:
$f\star(g\star h)=(f\star g)\star h$,
and returns back to the ordinary product
in the commutative limit:  $\theta^{ij}\rar 0$.
The modification of the product  makes the ordinary
spatial coordinate ``noncommutative,''
that is, $[x^i,x^j]_\star:=x^i\star x^j-x^j\star x^i=i\theta^{ij}$.

We note that the fields themselves take c-number values
as usual and the differentiation and the integration for them
are well-defined as usual.
A nontrivial point is that
NC field equations contain infinite number of
derivatives in general. Hence the integrability of the equations
are not so trivial as commutative cases.

\subsection{ADHM Construction of Instantons}

In this subsection, we treat NC instantons by
Atiyah-Drinfeld-Hitchin-Manin (ADHM) construction \cite{ADHM}.
ADHM construction is a strong method to generate instanton
solutions with arbitrary instanton number for $SU(N), SO(N)$ and
$Sp(N)$. This is based on a duality, that is, one-to-one
correspondence between the instanton moduli space and the moduli
space of ADHM-data which are specified by the ASD equation and
ADHM equation, respectively. The concrete steps are as follows
(For reviews on commutative spaces, see e.g
\cite{DHKM,Hamanaka,Hamanaka3}.):
\begin{itemize}

\item Step (i):  Solving ADHM equation:
\begin{eqnarray}
\label{adhm_now}
&&{[B_1,B_1^\dagger]}+[B_2,B_2^\dagger]+II^\dagger-J^\dagger J
=-[z_1,\zb_1]-[z_2,\zb_2]=0,\nonumber\\
&&{[B_1,B_2]}+IJ=-[z_1,z_2]=0.
\end{eqnarray}
We note that the coordinates $z_{1,2}$ always appear in pair with
the matrices $B_{1,2}$ and that is why we see the commutator of
the coordinates in the RHS.
These terms, of course, vanish on commutative spaces, however,
they cause nontrivial contributions on NC spaces,
which is seen later soon.

\item Step (ii):  Solving ``0-dimensional Dirac equation'' in the
      background of the ADHM date:
\begin{eqnarray}
\label{0dirac_now}
\na^\dagger V=0,
\end{eqnarray}
with the normalization condition:
\begin{eqnarray}
\label{0norm_now}
V^\dagger V =1.
\end{eqnarray}

\item Step (iii): By using the solution $V$,
we can construct the corresponding instanton solution as
\begin{eqnarray}
\label{4inst_now}
A_\mu=V^\dagger \del_\mu V,
\end{eqnarray}
which actually satisfies the ASD equation:
\begin{eqnarray}
\label{asd_now}
&&F_{z_1\zb_1}+F_{z_2\zb_2}=
[D_{z_1},D_{\zb_1}]+[D_{z_2},D_{\zb_2}]=0,\nn
&&F_{z_1z_2}={[D_{z_1},D_{z_2}]}=0.
\end{eqnarray}

\end{itemize}

In this subsection, we give some examples of
the explicit instanton solutions
focusing on Belabin-Polyakov-Schwartz-Tyupkin (BPST) 
instanton solutions.

\vs
\noindent
\unl{{\it BPST instanton solution}
(1-instanton, $G=SU(2)$)}
\vs

This solution is the most basic and important
and is constructed almost trivially
by ADHM procedure.

\begin{itemize}
\item Step (i): ADHM equation is a $k\ti k$ matrix-equation
and in the present $k=1$ case, it is trivially solved.
The commutator part of $B_{1,2}$ is automatically dropped out
and the matrices $B_{1,2}$ can be taken as arbitrary complex
numbers.
The remaining part $I, J$ are also easily solved:
\begin{eqnarray}
B_1=\alpha_1,~B_2=\alpha_2,~I=(\rho,0),~ J=\left(
\begin{array}{c}
0\\
\rho
\end{array}
\right),~~~\alpha_{1, 2}\in \C,~\rho\in \R.
\end{eqnarray}
Here the real and imaginary part of
$\alpha$ are denoted by $\alpha_1=b_2+ib_1,~\alpha_2=b_4+ib_3$,
respectively.

\item Step (ii): The ``0-dimensional Dirac equation''
is also easily solved in this case. (See, e.g. \cite{Hamanaka3}.)

\item Step (iii): The instanton solution is constructed as follows
\begin{eqnarray}
\lab{bpstkai}
A_\mu=V^\dagger \del_\mu V
&=&\fr{i(x-b)^\nu\eta^{(-)}_\mn }{(x-b)^2+\rho^2}.
\end{eqnarray}
The field strength $F_\mn$ is calculated from this gauge field as
\begin{eqnarray}
F_\mn=\fr{2i\rho^2}{(\vert z-\alpha\vert^2+\rho^2)^2}\eta_{\mn}^{(-)}.
\end{eqnarray}
This is just the BPST instanton solution \cite{BPST}.
The distribution is just like in Fig. \ref{bpst_com}.
The dimension 5 of the instanton moduli space
corresponds to the positions $b^\mu$ and the size $\rho$
of the instanton.

Now let us take the zero-size limit. Then the distribution of the
field strength $F_{\mn}$ converses into a singular,
delta-functional configuration. Instantons have smooth
configurations by definition and hence the zero-size instanton
does not exists, which corresponds to the singularity of the
(complete) instanton moduli space which is called the {\it small
instanton singularity}. (See Fig. \ref{bpst_com}.)
On NC spaces, the singularity is resolved and new class
of instantons appear.

\begin{figure}[htbn]
\epsfxsize=90mm
\hspace{2cm}
\epsffile{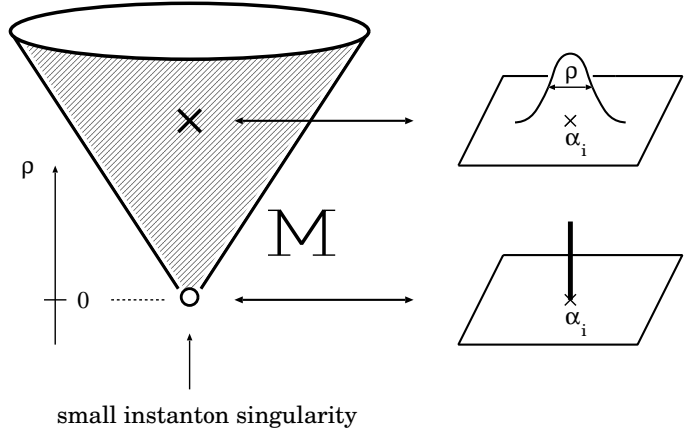}
\caption{Instanton moduli space $M $ and
the instanton configurations
(The horizontal directions correspond to the degree of global gauge
transformations which act on the gauge fields as the adjoint
action.) }
\label{bpst_com}
\end{figure}

\end{itemize}

\subsection{ADHM Construction of NC Instantons}

In this subsection, we construct some typical NC instanton
solutions by using ADHM method in the operator formalism. In NC
ADHM construction, the self-duality of the NC parameter is
important, which reflects the properties of the instanton
solutions.

The steps are all the same as the commutative one:
\begin{itemize}

\item Step (i): ADHM equation is deformed by the noncommutativity of
the coordinates as we mentioned in the previous subsection:
\begin{eqnarray}
\label{nc_adhm}
(\mu_{\scr\mbox{\R}}:=)
&&{[B_1,B_1^\dagger]}+[B_2,B_2^\dagger]+II^\dagger-J^\dagger J
=-2(\theta_{1}+\theta_{2})=:\zeta,\nonumber\\
(\mu_{\scr\mbox{\C}}:=)&&{[B_1,B_2]}+IJ=0.
\end{eqnarray}
We note that if the NC parameter is ASD, that is,
$\theta_1+\theta_2=0$, then the RHS of the first equation of ADHM
equation becomes zero.\footnote{When we treat SD gauge fields,
then the RHS is proportional to $(\theta_1-\theta_2)$. Hence the
relative self-duality between gauge fields and NC parameters is
important.}

\item Step (ii): Solving the NC ``0-dimensional Dirac
      equation''
\begin{eqnarray}
\label{nc_0dirac}
\hat{\na}^\dagger \Vh=0
\end{eqnarray}
with the normalization condition.

\item Step (iii): the ASD gauge fields are
constructed from the zero-mode $V$,
\begin{eqnarray}
\label{nc_4inst}
\Ah_\mu=\Vh^\dagger \del_\mu \Vh,
\end{eqnarray}
which actually satisfies the NC ASD equation:
\begin{eqnarray}
\label{nc_asd}
(\Fh_{z_1\zb_1}+\Fh_{z_2\zb_2}=)&&
[\Dh_{z_1},\Dh_{\zb_1}]+[\Dh_{z_2},\Dh_{\zb_2}]-\half\left(\fr{1}{\theta_1}
+\fr{1}{\theta_2}\right)=0,\nn
(\Fh_{z_1z_2}=)&&{[\Dh_{z_1},\Dh_{z_2}]}=0.
\end{eqnarray}
There is seen to be a beautiful duality between (\ref{nc_adhm})
and (\ref{nc_asd}) We note that when the NC parameter is ASD, then
the constant terms in both (\ref{nc_adhm}) and (\ref{nc_asd})
disappear.

\end{itemize}

In this way, NC instantons are actually constructed. 
We note that the gauge group is not $SU(N)$ but $U(N)$
because  $g_1,g_2\in SU(N)$ does not neccessarily 
lead to $g_1\star g_2 \in SU(N)$. This $U(1)$ part 
of the gauge group play important roles.

\vs
\noindent
\underline{\bf Comments on instanton moduli spaces}
\vs

\begin{figure}[htbn]
\epsfxsize=80mm
\hspace{2cm}
\epsffile{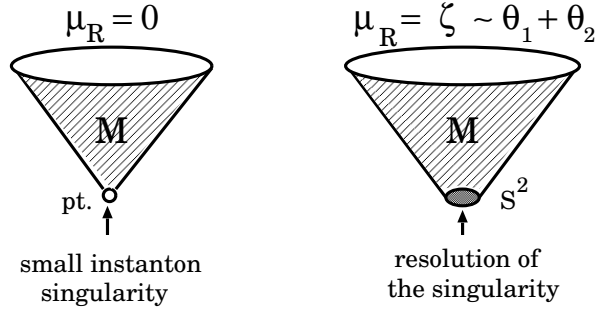}
\caption{Instanton Moduli Spaces}
\label{moduli}
\end{figure}

Instanton moduli spaces are determined by the value of
$\mu_{\scr\mbox{\R}}$
\cite{Nakajima2,Nakajima3}. (cf. Fig. \ref{moduli}.)
Namely,
\begin{itemize}
\item  In $\mu_{\scr\mbox{\R}}=0$ case, instanton moduli spaces
contain small instanton singularities, (which is the case for
commutative $\R^4$ and special NC $\R^4$ where
 $\theta$ : ASD).
\item In $\mu_{\scr\mbox{\R}}\neq 0$ case,
small instanton singularities are resolved and new class of smooth
instantons, U(1) instantons exist, (which is the case for general
NC $\R^4$)
\end{itemize}

Since $\mu_{\scr\mbox{\R}}=\zeta=-2(\theta_1+\theta_2)$ as Eq.
(\ref{nc_adhm}), the self-duality of the NC parameter is
important. NC ASD instantons have the following ``phase diagram''
(Fig. \ref{phase}):

\begin{figure}[htbn]
\epsfxsize=60mm
\hspace{3cm}
\epsffile{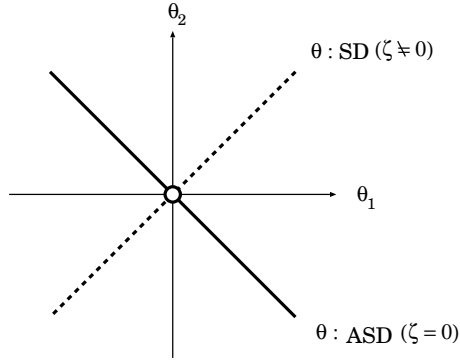}
\caption{``phase diagram'' of NC ASD instantons}
\label{phase}
\end{figure}

When the NC parameter is ASD, that is, $\theta_1+\theta_2=0$,
instanton moduli space implies the singularities. The origin of
the ``phase diagram'' corresponds to commutative instantons. The
$\theta$-axis represents instantons on $\R^2_{\scr{\mbox{NC}}}\ti
\R^2_{\scr{\mbox{Com}}}$. The other instantons basically have the
same properties, hence let us fix the NC parameter $\theta$
self-dual. This type of instantons are just discussed first by
Nekrasov and Schwarz \cite{NeSc}. The ASD-SD instantons (the
combination of self-dualities of gauge fields and NC
parameters is ASD-SD) are discussed in e.g.
\cite{CLMS,FSIv,Furuuchi,Furuuchi2,Furuuchi3,Ho,IKS,ILME,KLY,KLY3,LePo,LeYa,LTY,Nekrasov,NeSc,Parvizi,Tian,TiZh}.
The ASD-ASD instantons \cite{AGMS} are constructed by ADHM
construction in \cite{Furuuchi4,Hamanaka2,Wimmer}, and ADHM
construction of instantons on $\R^2_{\scr{\mbox{NC}}} \ti
\R^2_{\scr{\mbox{Com}}}$ are discussed in \cite{KLY2}.
Witten's ansatz \cite{Witten}
for NC instantons are studied in \cite{CFMSS,CMS,Schaposnik}. 
Geometrical origin of instanton number of NC instantons is also discussed in
e.g. \cite{Furuuchi3,Harvey3,HaMo,IKS2,Matsuo,Sako,Schwarz2,Tian2,STZ}.
For comprehensive discussion on ADHM construction, 
see e.g. \cite{CKT,Hamanaka3,Watamura}.
Instantons in Born-Infeld actions in the background of B-fields
are discussed in \cite{HaOo,KrSh,Moriyama2,SeWi,Terashima}.

\vs
\noindent
\unl{{\it NC BPST instanton solution}
(1-instanton, $G=U(2)$, $\theta$: SD)}
\vs

This solution is also obtained by ADHM procedure with the
``Furuuchi's Method'' \cite{Furuuchi,Furuuchi2}.  The solution of
NC ADHM equation is
\begin{eqnarray}
B_{1,2}=0,~~~I=(\sqrt{\rho^2+\zeta},0),~~~J=\left(\begin{array}{c}0\\\rho\end{array}\right).
\end{eqnarray}
Comparing the solution of commutative ADHM equation, the date $I$
is deformed by the noncommutativity of the coordinates, which
shows that the size of instantons becomes larger than that of
commutative one because of the existence of $\zeta$. In fact, in
the $\rho\rar 0$ limit, the configuration is still smooth and the
U(1) part is alive. This is essentially the same as a
$U(1)$ instanton solution.

BPST instantons on commutative and NC spaces are summarized as
follows.

\begin{center}
\begin{tabular}{|c|c|c|} \hline
BPST instanton& &NC BPST instanton\\ \hline\hline
$\mu_{\scr\mbox{\bf R}}=0,~\mu_{\scr\mbox{\bf C}}=0$&
ADHM equation&$\mu_{\scr\mbox{\bf R}}=\zeta ,~\mu_{\scr\mbox{\bf C}}=0$
\\\hline
$B_{1,2}=\alpha_{1,2},$&ADHM data
&$B_{1,2}=\alpha_{1,2},$\\
$I=(\rho,0),J^t=(0,\rho)$&~&$I=(\sqrt{\rho^2+\zeta},0),J^t=(0,\rho)$
 \\\hline
{\bf R}$^4\times$ orbifold ${\bf C}^2/{\bf Z}_2$& moduli
space&{\bf R}$^4\times$ Eguchi-Hanson $\widetilde{{\bf C}^2/{\bf Z}_2}$\\
(singular)&~&(regular) \\\hline
$F_{\mn}\rar$ delta function&zero-size limit
& $F_{\mn}\rar$ $U(1)$ instanton\\
~~~~~~~(singular)&~&~~(regular) \\\hline
\end{tabular}
\end{center}

More detailed discussion are presented in e.g.
\cite{Furuuchi3,Hamanaka3,KoSc2,Lechtenfeld,Nekrasov2,Schaposnik,Watamura}.

\vs \noindent
\underline{\bf Some other BPS solitons}
\vs

There are many works on the study of other NC BPS solitons as
follows:
\begin{itemize}
\item NC monopoles:
\cite{Bak,GoMa,GrNe,GrNe2,GrNe3,Hamanaka2,HaTe,Hashimoto,Hashimoto2,HaHi,HHM,HHM2,Moriyama,Nekrasov2,Polychronakos,PSW}
\item NC vortices in abelian Higgs models:
\cite{Bak2,BLP,JMW,LMS}
\item NC solitons in $CP(n)$ models:
\cite{FJJ,FINY,Ghosh,GoHa,LLY,AdMu,IKOS}
\item Higher-dim. NC instantons:
\cite{FIO,HIO,Hiraoka,IvLe,KLY4,LPS2,MPT,Nekrasov3,Ohta_k,PoSz,SaSu,Valtancoli,Witten2}
\end{itemize}

\section{Towards NC Sato's Theories}


\subsection{NC Gelfand-Dickey's Hierarchies}
In this section, we derive various NC soliton equations in terms
of pseudo-differential operators which include negative powers of
differential operators.

An $N$-th order (monic) pseudo-differential operator $A$ is
represented as follows
\begin{eqnarray}
 A=\del_x^N + a_{N-1}\del_x^{N-1}+ \cdots
+ a_0 +a_{-1}\del_x^{-1}+a_{-2}\del_x^{-2}+\cdots.
\end{eqnarray}
Here we introduce useful symbols:
\begin{eqnarray}
 A_{\geq r}&:=& \del_x^N + a_{N-1}\del_x^{N-1}+ \cdots + a_{r}\del_x^{r},\\
 A_{\leq r}&:=& A - A_{\geq r+1}
 = a_{r}\del_x^{r} + a_{r-1}\del_x^{r-1} +\cdots,\\
 {\mbox{res}}_{r} A &:=& a_{r}.
\end{eqnarray}
The symbol ${\mbox{res}}_{-1} A$ is especially called the {\it residue} of $A$.

The action of a differential operator $\partial_x^n$ on
a multiplicity operator $f$ is formally defined
as the following generalized Leibniz rule:
\begin{eqnarray}
 \partial_x^{n}\cdot f:=\sum_{i\geq 0}
\left(\begin{array}{c}n\\i\end{array}\right)
(\partial_x^i f)\partial_x^{n-i},
\end{eqnarray}
where the binomial coefficient is given by
\begin{eqnarray}
\label{binomial}
 \left(\begin{array}{c}n\\i\end{array}\right):=
\frac{n(n-1)\cdots (n-i+1)}{i(i-1)\cdots 1}.
\end{eqnarray}
We note that the definition of the binomial coefficient (\ref{binomial})
is applicable to the case for negative $n$,
which just define the action of
negative power of differential operators.
The examples are,
\begin{eqnarray}
 \partial_x^{-1}\cdot f&=&
f\partial_x^{-1}-f^\prime\partial_x^{-2}
+f^{\prime\prime}\partial_x^{-3}-\cdots,\nn
 \partial_x^{-2}\cdot f&=&
f\partial_x^{-2}-2f^\prime\partial_x^{-3}
+3f^{\prime\prime}\partial_x^{-4}-\cdots,\nn
 \partial_x^{-3}\cdot f&=&
f\partial_x^{-3}-3f^\prime\partial_x^{-4}
+6f^{\prime\prime}\partial_x^{-5}-\cdots,
\end{eqnarray}
where $f^\prime:=\del f/\del x,
f^{\prime\prime}:=\del^2 f/\del x^2$ and so on,
and $\partial_x^{-1}$ in the RHS
acts as an integration operator $\int^x dx$.

The composition of pseudo-differential operators is also
well-defined and the total set of pseudo-differential operators
forms an operator algebra. For more on pseudo-differential
operators and Sato's theory, see e.g.
\cite{BBT,Blaszak,DJKM,Dickey,DJM,OSTT}.

\vspace{3mm}

Let us introduce a Lax operator
as the following first-order pseudo-differential operator:
\begin{eqnarray}
 L = \partial_x + u_2 \partial_x^{-1}
 + u_3 \partial_x^{-2} + u_4 \partial_x^{-3} + \cdots,
\end{eqnarray}
where the coefficients $u_k$ ($k=2,3,\ldots$) are functions
of infinite variables $(x^1,x^2,\ldots)$ with $x^1\equiv x$:
\begin{eqnarray}
 u_k=u_k(x^1,x^2,\ldots).
\end{eqnarray}
The noncommutativity is arbitrarily introduced for
the variables $(x^1,x^2,\ldots)$ as Eq. (\ref{nc_coord}) here.

The NC KP hierarchy is defined in Sato's framework as
\begin{eqnarray}
 \del_m L = \left[B_m, L\right]_\star,~~~m=1,2,\ldots,
\label{lax_sato}
\end{eqnarray}
where the action of $\del_m$ on the pseudo-differential operator $L$
should be interpreted to be coefficient-wise,
that is, $\del_m L :=[\del_m,L]$ or $\del_m \del_x^k=0$.
The operator $B_m$ is given by
\begin{eqnarray}
 B_m
:=(\underbrace{L\star \cdots \star L}_{ m{\scriptsize\mbox{
times}}})_{\geq 0}=:(L^m)_{\geq 0}.
\end{eqnarray}
The KP hierarchy gives rise to a set of infinite differential
equations with respect to infinite kind of fields from the
coefficients in Eq. (\ref{lax_sato}) for a fixed $m$. Hence it
contains huge amount of differential (evolution) equations for all
$m$. The LHS of Eq. (\ref{lax_sato}) becomes $\del_m u_k$ which
shows a flow in the $x^m$ direction.

If we put the constraint $L^l=B_l$ on the NC KP hierarchy
(\ref{lax_sato}), we get infinite set of $l$-reduced NC KP
hierarchies. We can easily show
\begin{eqnarray}
\label{Nl}
\frac{\partial u_k}{\partial x^{Nl}}=0,
\end{eqnarray}
for all $N,k$ because
\begin{eqnarray}
 \fr{dL^l}{dx^{Nl}}=[B_{Nl},L^l]_\star=[(L^{l})^N,L^l]_\star=0,
\end{eqnarray}
which implies Eq. (\ref{Nl}). The reduced NC KP hierarchy is
called the {\it l-reduction} of the NC KP hierarchy. This time,
the constraint $L^l=B_l$ gives simple relationships which make it
possible to represent infinite kind of fields
$u_{l+1},u_{l+2},u_{l+3},\ldots$ in terms of $(l-1)$ kind of
fields $u_{2},u_{3},\ldots, u_{l}$. (cf. Appendix A in
\cite{Hamanaka4}.)

{}Let us see explicit examples.
\begin{itemize}

\item NC KP hierarchy

The coefficients of each powers of (pseudo-)differential operators
in the NC KP hierarchy (\ref{lax_sato}) yield a series of infinite
NC ``evolution equations,'' that is, for $m=1$
\begin{eqnarray}
\partial_x^{1-k})~~~ \del _1 u_{k}=u_{k}^\prime,~~~k=2,3,\ldots
~~~\Rightarrow~~~x^1\equiv x,
\end{eqnarray}
for $m=2$
\begin{eqnarray}
\label{KP_hie}
\partial_x^{-1})~~~\del_2 u_{2}
&=&u_2^{\prime\prime}+2u_{3}^{\prime},\nonumber \\
\partial_x^{-2})~~~
\del_2 u_{3}&=&u_3^{\prime\prime}+2u_4^{\prime}
+2u_2\star u_2^\prime +2[u_2,u_3]_\star,\nonumber \\
\partial_x^{-3})~~~
\del_2 u_{4}&=&u_{4}^{\prime\prime}+2u_{5}^{\prime}
+4u_3\star u_2^\prime-2u_2\star u_2^{\prime\prime}
+2[u_2,u_4]_\star,\nn
\partial_x^{-4})~~~\del_2 u_{5}&=&\cdots,
\end{eqnarray}
and for $m=3$
\begin{eqnarray}
\label{3flow}
\partial_x^{-1})~~~
\del_3 u_{2}&=&u_{2}^{\prime\prime\prime}+3u_3^{\prime\prime}
+3u_4^{\prime}+3u_2^\prime\star u_2+3u_2\star u_2^\prime,
\nonumber\\
\partial_x^{-2})~~~
\del_3 u_{3}&=&u_{3}^{\prime\prime\prime}+3u_{4}^{\prime\prime}
+3u_{5}^\prime+6u_{2}\star u_{3}^\prime+3u_2^\prime\star u_3
+3u_3\star u_2^\prime+3[u_2, u_4]_\star,\nn
\partial_x^{-3})~~~
\del_3 u_{4}&=&u_{4}^{\prime\prime\prime}+3u_{5}^{\prime\prime}
+3u_{6}^\prime+3u_{2}^\prime \star u_{4}+3u_2\star u_4^\prime
+6u_4\star u_2^\prime\nn
&&-3u_2\star u_3^{\prime\prime}
-3u_3\star u_2^{\prime\prime}+6u_3\star u_3^{\prime}
+3[u_2,u_5]_\star+3[u_3,u_4]_\star,\nn
\partial_x^{-4})~~~\del_3 u_{5}&=&\cdots.
\end{eqnarray}
These just imply the $(2+1)$-dimensional NC KP equation
\cite{Paniak,Kupershmidt} with $2u_2\equiv u, x^2\equiv
y,x^3\equiv t$ and $\partial_x^{-1}f(x)=\int^x dx^\prime f(x^\prime)$:
\begin{eqnarray}
 \fr{\del u}{\del t}=\frac{1}{4}\fr{\del^3 u}{\del x^3}
+\frac{3}{4}\fr{\del (u\star u)}{\del x}
+\frac{3}{4}\partial_x^{-1} \fr{\del^2 u}{\del y^2}
-\frac{3}{4}\left[u,\partial_x^{-1} \fr{\del u}{\del
y}\right]_\star.
\end{eqnarray}
Important point is that infinite kind of fields $u_3, u_4, u_5,\ldots$
are represented in terms of one kind of field  $2u_2\equiv u$
as is seen in Eq. (\ref{KP_hie}).
This guarantees the existence of NC KP hierarchy
which implies the existence of reductions of
the NC KP hierarchy.
The order of nonlinear terms are determined in this way.

\item NC KdV Hierarchy (2-reduction of the NC KP hierarchy)

Taking the constraint $L^2=B_2=:\del_x^2+u$ for
the NC KP hierarchy, we get the NC KdV hierarchy.
This time, the following NC Lax hierarchy
\begin{eqnarray}
\label{KdV_hie}
 \frac{\partial u}{\partial x^m}=\left[B_m, L^2\right]_\star,
\end{eqnarray}
include neither positive nor negative power of
(pseudo-)differential operators for the same reason as commutative
case and gives rise to the $m$-th KdV equation for each $m$. For
example, the NC KdV hierarchy (\ref{KdV_hie}) becomes the
$(1+1)$-dimensional NC KdV equation \cite{DiMH3} for $m=3$ with
$x^3\equiv t$
\begin{eqnarray}
\label{ncKdV}
 \dot{u}=\frac{1}{4}u^{\prime\prime\prime}+\frac{3}{4}
\left(u^\prime \star u + u \star u^\prime \right),
\end{eqnarray}
and the $(1+1)$-dimensional 5-th NC KdV equation \cite{Toda}
for $m=5$ with $x^5\equiv t$
\begin{eqnarray}
\dot{u}&=&\frac{1}{16}u^{\prime\prime\prime\prime\prime}
+\frac{5}{16}(u\star
u^{\prime\prime\prime}+u^{\prime\prime\prime}\star u)
+\frac{5}{8}(u^{\prime}\star u^{\prime}+u\star u\star u)^\prime.
\end{eqnarray}

\item NC Boussinesq Hierarchy (3-reduction of the NC KP hierarchy)

The 3-reduction  $L^3=B_3$ yields the NC Boussinesq hierarchy
which includes the $(1+1)$-dimensional
NC Boussinesq equation \cite{Toda}
with $t\equiv x^2$:
\begin{eqnarray}
 \ddot{u}=\frac{1}{3}u^{\prime\prime\prime\prime}
+(u \star u)^{\prime\prime}
+([u,\partial_x^{-1}\dot{u}]_{\star})^\prime,
\end{eqnarray}
where $\ddot{u}:=\del^2 u/\del t^2$.
\end{itemize}

In this way, we can generate infinite set of the $l$-reduced NC KP
hierarchies. (This is called the {\it NC Gelfand-Dickey
hierarchies} which reduce to the ordinary Gelfand-Dickey
hierarchies \cite{GeDi} in the commutative limit.) The present
discussion is also applicable to the matrix Sato theory where the
fields $u_k$ ($k=1,2,\ldots$) are $N\times N$ matrices. For $N=2$,
the Lax hierarchy includes the Ablowitz-Kaup-Newell-Segur (AKNS)
system \cite{AKNS}, the Davey-Stewarson equation, the NLS equation
and so on. (For commutative discussions, see e.g.
\cite{Blaszak,Dickey}.) NC Bogoyavlenskii-Calogero-Schiff (BCS)
equation \cite{Toda} is also derived.

\subsection{Conservation Laws}

Here we prove the existence of infinite conservation laws for the
wide class of NC soliton equations. The existence of infinite
number of conserved quantities would lead to infinite-dimensional
hidden symmetry from Noether's theorem.

First we would like to comment on conservation laws
of NC field equations \cite{HaTo2}.
The discussion is basically the same as commutative case
because both the differentiation and the integration
are the same as commutative ones in the Moyal representation.

Let us suppose the conservation law
\begin{eqnarray}
\partial_t \sigma(t,x^i)=\partial_i J^i(t,x^i),
\end{eqnarray}
where $\sigma(t,x^i)$ and $J^i(t,x^i)$ are called
the {\it conserved density} and the {\it associated flux},
respectively.
The conserved quantity is given by spatial integral
of the conserved density:
\begin{eqnarray}
Q(t)=\int_{\scr\mbox{space}}d^Dx \sigma(t,x^i),
\end{eqnarray}
where the integral $\int_{\scr\mbox{space}}dx^D$
is taken for spatial coordinates.
The proof is straightforward:
\begin{eqnarray}
{d Q\over{dt}}&=&{\partial \over{\partial
t}}\int_{\scr\mbox{space}} d^Dx \sigma(t,x^i)
=\int_{\scr\mbox{space}} d^Dx \partial_i J_i(t,x^i)\nn
&=&\int_{\st{\scr\mbox{spatial}}{\scr\mbox{infinity}}}dS^i
J_i(t,x^i) =0,
\end{eqnarray}
unless the surface term of the integrand  $J_i(t,x^i)$ vanishes.
The convergence of the integral is also expected because
the star-product naively reduces to the ordinary product
at spatial infinity due to: $\del_i \sim {O}(r^{-1})$
where $r:=\vert x \vert$.

Here let us return back to NC hierarchy. In order to discuss the
conservation laws, we have to specify for what equations the
conservation laws are. The specified equations possess space and
time coordinates in the infinite coordinates $x_1,x_2,x_3,\cdots$.
Identifying $t\equiv x^m$, we can get infinite
conserved densities for the NC Lax hierarchies 
as follows ($n=1,2,\ldots$) \cite{Hamanaka4}:
\begin{eqnarray}
 \sigma_n={\mbox{res}}_{-1} L^n+\theta^{im}\sum_{k=0}^{m-1}\sum_{l=0}^{k}
\left(\begin{array}{c}k\\l\end{array}\right) \del_x^{k-l}{\mbox{res}}_{-(l+1)} L^n
\diamond \del_i {\mbox{res}}_{k} L^m,
\label{conservation}
\end{eqnarray}
where the suffices $i$ must run in the space-time directions only.
The symbol ``$\diamond$'' is called the {\it Strachan product}
\cite{Strachan} and defined by
\begin{eqnarray}
 f(x)\diamond g(x)
:=\sum_{s=0}^{\infty}
\fr{(-1)^s}{(2s+1)!}\left(\frac{1}{2}\theta^{ij}
\del_i^{(x^\prime)}\del_j^{(x^{\prime\prime})}\right)^{2s}
f(x^\prime)g(x^{\prime\prime})\vvert_{x^\prime=x^{\prime\prime}=x}.
\end{eqnarray}
This is a commutative and non-associative product.

We can easily see that deformation terms appear in the second term
of Eq. (\ref{conservation}) in the case of space-time
noncommutativity. On the other hand, in the case of space-space
noncommutativity, the conserved density is given by the residue of
$L^n$ as commutative case.

For examples, explicit representation of the NC KP equation with
space-time noncommutativity, the NC KdV equation is
\begin{eqnarray}
\sigma_n
={\mbox{res}}_{-1} L^n
-3\theta
\left(({\mbox{res}}_{-1}L^n)\diamond u_3^\prime
+({\mbox{res}}_{-2}L^n)\diamond u_2^\prime
\right).
\end{eqnarray}

We have a comment on conserved densities
for one-soliton configuration.
One soliton solutions can be always reduced to commutative ones
as is seen in the next subsection.
Hence the conserved densities are not deformed
in the NC extension.

The present discussion is applicable to
the NC matrix Sato theory,
including the NC AKNS system,
the NC Davey-Stewarson equation,
the NC NLS equation, and the NC BCS equation.

\subsection{Some Exact Solutions}

Here we show the existence of exact (multi-soliton) solutions by
giving the explicit formula.

First, let us comment on one-soliton solutions \cite{DiMH3,HaTo2}.
Defining $z:=x+vt,
\zb:=x-vt$, we easily see
\begin{eqnarray}
 f(z)\star g(z)= f(z) g(z)
\end{eqnarray}
because the star-product (\ref{star}) is rewritten in terms of
$(z,\zb)$ as
\begin{eqnarray}
 f(z,\zb)\star g(z,\zb)=
e^{iv\theta\left(
\partial_{\zb^\prime}
\partial_{z^{\prime\prime}}-
\partial_{z^\prime}
\partial_{\zb^{\prime\prime}}
\right)}f(z^\prime,\zb^\prime)
g(z^{\prime\prime},\zb^{\prime\prime}) \Big{\vert}_{\scr
\begin{array}{c} z^{\prime}
=z^{\prime\prime}=z\\
\zb^{\prime} =\zb^{\prime\prime}=\zb. \end{array}}
\end{eqnarray}
Hence NC one soliton-solutions are essentially the same as 
commutative ones.

Next, we prove that NC Burgers equations derived from NC
Gelfand-Dickey hierarchies are integrable in the sense that they
are linearizable.

NC Burgers equation is obtained by a special reduction of 
NC mKP hierarchies \cite{HaTo2,Hamanaka4}:
\begin{eqnarray}
 \dot{u}-u^{\prime\prime}-2u\star u^{\prime}=0.
\label{burgers}
\end{eqnarray}
The solutions of the following NC diffusion equations
\begin{eqnarray}
\label{diffusion} \dot{\psi}=\psi^{\prime\prime},
\end{eqnarray}
solve Eq.(\ref{burgers}) via the NC Cole-Hopf transformation:
$u=\psi^{-1}\star \psi^\prime$. The naive solution of the NC
diffusion equation (\ref{diffusion}) is
\begin{eqnarray}
\label{naive_sol} \psi(t,x)=1+\sum_{i=1}^N h_i e^{a k_i^2t}\star
e^{\pm k_i x} =1+\sum_{i=1}^N h_i e^{\fr{i}{2}a k_i^3\theta}e^{a
k_i^2t\pm k_i x},
\end{eqnarray}
where $h_i, k_i$ are complex constants. The final form in
(\ref{naive_sol}) shows that the naive solution on commutative
space is deformed by $e^{\fr{i}{2}a k_i^3\theta}$ due to the
noncommutativity. This reduces to the $N$-shock wave solution in
fluid dynamics. Hence we call it the {\it NC N-shock wave
solution}. Exact solutions for $N=1,2$ are obtained by L.~Martina
and O.~Pashaev \cite{MaPa} in terms of $u$ and nontrivial effects
of the NC-deformation are actually reported.

This is a very interesting result. The NC Burgers equation
contains infinite number of time derivatives in the nonlinear term
and integrability would be naively never expected. Initial value
problems are hard to define. Nevertheless, the NC Burgers equation
is linearizable and the linearized equation is a differential
equation of first order with respect to time and the initial value
problem is well-defined. This shows that the NC Burgers equation
is completely integrable.

General arguments for NC hierarchies are possible. Exact solutions
for them are already given by Etingof, Gelfand and Retakh
\cite{EGR} as explicit forms in terms of quasi-determinants
\cite{GeRe}. (For a survey of quasi-determants, see \cite{GGRW}.)
In Moyal deformations, the solutions are actually
{\it multi-soliton} solutions,
in the sense that
the configuration has localized energy.
This can be seen in the asymptotic
behavior. In scattering processes, the soliton configurations
are stable and never decay. Noncommutativity affects 
the phase shifts only. Exact solutions for NC KP eq. 
would coincide with those by Paniak
\cite{Paniak}. More detailed discussion will be reported later
soon \cite{Hamanaka5}. Exact solutions are also discussed in
\cite{DiMH6,Sakakibara,WaWa}.

\subsection{Some Examples of NC Ward's Conjecture}

In this subsection, we present some examples of NC Ward's
conjecture, including NC NLS eq., NC KdV eq.,
NC Burgers eq. and so on. (For commutative discussions, see e.g.
\cite{ACH,ACT,AbCl,CKN,IvPo,MaWo}.)

\begin{itemize}

\item NC NLS and KdV equations

Let us consider the following NC ASDYM equation with $G=U(2)$,
which is dimensionally reduced to 2-dimensional space-time (The
convention is the same as \cite{MaWo}. See also  \cite{Hamanaka6}.):
\begin{eqnarray}
 &&Q^\prime=0,~~~
 \dot{Q} + \Phi_{w}^\prime
 +[\Phi_{z},Q]_\star =0,\nn
 &&\Phi_{z}^\prime -\dot{\Phi}_{w}
 +[\Phi_{w},\Phi_{z}]_\star=0.
\label{asdym}
\end{eqnarray}
where $Q, \Phi_{w}$ and $\Phi_{z}$ denote the original gauge
fields.

Now let us take a further reduction
on the gauge fields in the ASDYM eq. (\ref{asdym})
as follows \cite{Legare}:
\begin{eqnarray}
 Q=\frac{i}{2} \left(\begin{array}{cc}1&0\\0&-1\end{array}\right),~
 \Phi_{w}=\left(\begin{array}{cc}0&\psi\\-\bar{\psi}&0\end{array}\right),~
 \Phi_{z}=i\left(\begin{array}{cc}\psi\star\bar{\psi}&-\psi^\prime\\
-\bar{\psi}^\prime&-\bar{\psi}\star\psi
\end{array}\right).
\end{eqnarray}
Then the NC ASDYM (\ref{asdym}) reduces to
\begin{eqnarray}
 i\dot{\psi}=\psi^{\prime\prime}+2\psi \star \bar{\psi} \star \psi.
\end{eqnarray}
This is just the NC NLS equation \cite{DiMH2}.

We note that the gauge group is not $SU(2)$ but $U(2)$ on NC
spaces because the matrix $\Phi_{z}$ is not traceless. This is a
very consistent result because in the original NC Yang-Mills
theories, $U(1)$ part of the gauge group is essential
\cite{DoNe,Hamanaka3,Szabo}.

Now let us take another further reduction 
on the gauge fields in the ASDYM eq. (\ref{asdym})
as follows:
\begin{eqnarray}
&& Q=\left(\begin{array}{cc}0&0\\1&0\end{array}\right),~
 \Phi_{w}=\left(\begin{array}{cc}q&~
-1\\q^\prime+q\star q&~-q\end{array}\right),\\
&& \Phi_{z}=\left(\begin{array}{cc}
\displaystyle\frac{1}{2}q^{\prime\prime}+q^\prime\star q&-
q^\prime\\
 \displaystyle\frac{1}{4}q^{\prime\prime\prime}
+\frac{1}{2}q^{\prime}\star q^{\prime}
+\frac{1}{2}\left\{q, q^{\prime\prime}\right\}_\star
+q\star q^\prime \star q
&~-\displaystyle\frac{1}{2}q^{\prime\prime}-q\star q^\prime
\end{array}\right),\nonumber
\end{eqnarray}
which reduces to discussion by Mason and Sparling \cite{MaSp}
in the commutative limit.
We can see that solutions of the NC potential KdV (pKdV) equation
\begin{eqnarray*}
\dot{q}=\displaystyle\frac{1}{4}q^{\prime\prime\prime}
+\frac{3}{2}q^{\prime}\star q^{\prime},
\end{eqnarray*}
solve the reduced NC ASDYM equation (\ref{asdym}).
NC pKdV equation is derived from the NC KdV equation 
\begin{eqnarray}
 \dot{u}=\displaystyle\frac{1}{4}u^{\prime\prime\prime}
+\frac{3}{4}\left(u^{\prime}\star u+u\star u^{\prime}\right),
\label{kdv} 
\end{eqnarray}
by setting $2q^\prime =u$.
Hence NC KdV equation is actually derived from
NC ASDYM equation in this way.

We note that for both cases, 
the gauge group must not be 
$SL(2,{\bf R})$ but $GL(2,{\bf R})$
on NC spaces because the matrix $\Phi_{z}$ is not traceless.
Traceless case never leads to NC NLS and KdV equations \cite{Hamanaka6}.
This result reflects the importance of
$U(1)$ part of the original gauge group.
However, it is naively hard to correspond this reduction
to a D-brane configuration.

\item NC Burgers equation

Let us consider the following NC ASDYM
equation with $G=U(1)$ (Eq.
(3.1.2) in \cite{MaWo}):
\begin{eqnarray}
 &&\del_{w} A_z -\del_{z} A_w+[A_w,A_z]_\star =0,~~~
 \del_{\tilde{w}} A_{\tilde{z}} -\del_{\tilde{z}} A_{\tilde{w}}
 +[A_{\tilde{w}},A_{\tilde{z}}]_\star =0,\nn
 &&\del_{z} A_{\tilde{z}} -\del_{\tilde{z}} A_{z}
 +\del_{\tilde{w}} A_{w} -\del_{w} A_{\tilde{w}}
 +[A_z,A_{\tilde{z}}]_\star
 -[A_{w},A_{\tilde{w}}]_\star=0.
\label{asdym2}
\end{eqnarray}
where $(z,\tilde{z},w,\tilde{w})$ and $A_{z,\tilde{z},w,\tilde{w}}$
denote the coordinates
of the original $(2+2)$-dimensional space and
the gauge fields, respectively.
We note that the commutator part should remain
though the gauge group is $U(1)$ because the elements of the
gauge group could be operators and the gauge group could be
considered to be non-abelian: $U(\infty)$.
This commutator part actually plays an important role
as usual in NC theories.

Now let us take the simple dimensional reduction $\del_{\tilde{z}}
=\del_{\tilde{w}}=0$ and put the following constraints (with
$w\equiv t,~z\equiv x$):
\begin{eqnarray}
 A_{\tilde{z}}=A_{\tilde{w}}=0,~~~A_z=u,~~~A_w= u^\prime+u\star u.
\end{eqnarray}
Then the NC ASDYM equation (\ref{asdym2})
reduces to :
\begin{eqnarray}
 \dot{u}-u^{\prime\prime}-2u\star u^{\prime}=0.
\end{eqnarray}
This is just the NC Burgers equation which is linearizable and
hence completely integrable in this sense \cite{HaTo2,MaPa}. We
note that without the commutator part $[A_w,A_z]_\star$, the
nonlinear term should be symmetric: $u^{\prime}\star u + u\star
u^{\prime}$, which leads to neither the Lax representation nor
linearized equations via a NC Cole-Hopf transformation
\cite{HaTo2}. This shows that the special feature in the original
NC gauge theories plays a crucial role in integrability for the
lower-dimensional equation. Therefore the NC Burgers equation is
expected to have some non-trivial property special to NC spaces
such as the existence of $U(1)$ instantons.

\item NC KdV equation

Finally Let us consider another reduction of NC ASDYM eq. onto NC KdV eq.
which is different from both the previous reduction
and that by Legar\'e \cite{Legare}.
Let us start with the following NC
ASDYM equation with $G=SL(2,{\bf R})$,
which is dimensionally reduced to 2-dimensional space-time
(The convention is the same as \cite{BaDe}):
\begin{eqnarray}
 &&[P,B]=0,~~~
 P^\prime- Q^\prime
 +[P,Q]_\star +[H,B]_\star =0,\nn
 &&\dot{Q}-H^\prime +[Q,H]_\star=0.
\label{asdym3}
\end{eqnarray}
where $B,H,P$ and $Q$ denote the original gauge fields.

Now let us take further reduction
on the gauge fields as follows:
\begin{eqnarray}
 &&B=\left(\begin{array}{cc}0&0\\-1&0\end{array}\right),~~~
 P=\frac{1}{2}\left(\begin{array}{cc}0&0\\-u&0\end{array}\right),\nn
 &&Q=\left(\begin{array}{cc}0&1\\-u&0\end{array}\right),~~~
 H=\frac{1}{4}\left(\begin{array}{cc} -u^\prime&~ 2u \\
 -u^{\prime\prime}-2u\star u&~ u^\prime
\end{array}\right).
\end{eqnarray}
Then the NC ASDYM equation (\ref{asdym3}) reduces to
\begin{eqnarray}
  \dot{u} +\frac{1}{4} u^{\prime\prime\prime}+\frac{3}{4}
  \left(u^\prime \star u + u \star u^\prime \right)=0.
\end{eqnarray}
This is just the NC KdV equation (\ref{ncKdV}).

NC KP equation is also derived in a similar way \cite{Hamanaka6}. 
These results are new. We note that the gauge group is $SL(2,{\bf R})$,
this time.

\end{itemize}

In this way, we can derive various integrable equations from NC
ASDYM eqs. by reductions. Existence of these reductions guarantees
the lower-dimensional integrable equations actually have the
corresponding physical situations and could be applied to analysis
of D-brane dynamics in special reduced situations.

\vs \noindent \underline{\bf An (incomplete) list of works on NC
integrable equations}
\begin{itemize}
\item NC Burgers eqs:
\cite{HaTo2,MaPa}
\item NC Fordy-Kulish systems:
\cite{DiMH4}
\item NC KdV eqs:
\cite{DiMH3,EGR}
\item NC KP eqs:
\cite{EGR,Paniak,Kupershmidt,WaWa}
\item NC Non-Linear Schr\"odinger eqs:
\cite{DiMH2,WaWa2}
\item NC Liouville, sine-Gordon, sinh-Gordon and Toda field eqs:
\cite{BCR,Cabrera-Carnero,CCMo,DiMH,GMPT,GrPe,Lechtenfeld2,LMPPT,K.Lee,Zuevsky}
\item NC hierarchies etc.:
\cite{DiMH6,EGR,Hamanaka4,HaTo3,Sakakibara,WaWa3}
\item NC dressing and splitting methods:
\cite{Bieling,HLW,IhUh,LePo,LePo2,LePo3,LPS,Wolf}
\item NC mini-twistor spaces:
\cite{LePo3}
\item NC twistor theories:
\cite{Hannabuss,IhUh,KKO,Takasaki}

\end{itemize}

\section{Conclusion and Discussion}

In the present article, we reported recent developments of NC
extension of soliton theories and integrable systems focusing on
ADHM construction of NC instantons and NC Sato's theories. 
In the former part, we saw how ADHM constructions work
and the small instanton singularities are resolved on NC spaces.
In the latter part, we proved the existence of infinite number 
of conserved densities and exact multi-soliton solutions 
for wide class of NC soliton equations.
This suggests that NC soliton equations could be completely integrable 
in some sense and an infinite-dimensional symmetry would be hidden. 

As a next step, completion of NC Sato's theory
is the most worth keeping to investigate. 
In order to reveal what the hidden symmetry is, we have to
construct theories of tau-functions which play crucial roles in
Sato's theories. (See also \cite{DiMH6,WaWa2,Sakakibara}.)
The symmetry would be represented in terms of 
some kind of deformed infinite-dimensional affine Lie algebras. 

From the original motivation, confirmation of NC Ward's conjecture
would be the most important via the construction of 
NC twistor theories. Some aspects of NC Twistor theories 
have been already discussed by many authors e.g. 
\cite{Hannabuss,IhUh,KKO,LePo3,Takasaki}. This
would clarifies integrability of NC ASDYM equations.
 
In reductions of ASDYM equations, we mainly should 
take metric of $(2,2)$-type signature 
which is called the {\it split signature}.
ASDYM theories with the split signature
can be embedded \cite{LPS} in
$N=2$ string theories \cite{OoVa}.  
Simple reductions of them are studied intensively
by Lechtenfeld's group \cite{Bieling,IhUh,LePo,Wolf}.
This guarantees that NC integrable equations would 
have physical meanings and might lead to various successful 
applications to the corresponding D-brane
dynamics and so on. It is also very interesting to
clarify what symmetries in reductions guarantee
integrabilities of lower dimensional integrable
equations. One approach should be provided from the viewpoint of 
Lagrangian formalism in supersymmetric Yang-Mills theories.
The BPS equations would just correspond to integrable equations
and preserved supersymmetries would 
relate to their integrability \cite{Hamanaka6}.
Various BPS D-brane configurations (e.g. \cite{EINOOST}) 
might have a relation to our studies.

Supersymmetric extension (e.g \cite{Legare,NiRa})
and higher dimensional extension (e.g. \cite{Toda})
would be interesting and straightforwardly possible.
Extension to non(-anti)commutative superspaces is
also considerable. We also expect special properties
would still survive in these extensions.

For space-time noncommutativity, we have to
reconsider foundation of Hamiltonian
formalism from the beginning in order to 
establish what the integrability for them is, 
especially, symplectic structures, Poisson brackets,
Liouville's theorem, Noether's theorem,
action-angle variables, initial value problems
and so on. Geometrical interpretations of them 
must be also clarified. Then meaning of existence of
infinite (non-local) conserved densities would be clear.

There are still many things worth studying to be seen.

\section*{Acknowledgements}

It is a great pleasure to thank the organizers for invitation and
hospitality during the COE workshop on NC Geometry and
Physics, 26 February - 3 March, 2004 at Keio university, and the
workshop on NC Integrable Systems, 18 August, 2004 at
the Max Planck Institute (MPI) for Flow Research, G\"ottingen, and
the workshop on Differential Geometry in Nagoya, 18-21 December, 2004.
He would be grateful to H.~Awata, A.~Dimakis, S.~Kakei, M.~Kato,
I.~Kishimoto, O.~Lechtenfeld, K.~Lee, L.~Mason, M-y. Mo,
A.~Mukherjee, F.~M\"uller-Hoissen, A.~Nakamula, S.~Odake,
D.~Popov, T.~Suzuki, K.~Takasaki and K.~Toda for useful comments. Thanks are
also due to organizers and audiences during the workshops
YITP-W-03-07 on ``QFT 2003'' and YITP-W-04-03 on ``QFT 2004,'' 
and the XVII Workshop ``Beyond the Standard Models,''
at Bad Honnef,
for hospitality and discussion, and to L.~Mason, O.~Lechtenfeld,
F.~M\"uller-Hoissen and K.~Lee for financial support and
hospitality during the stays at Oxford university,
Hannover university, MPI for Flow Research, G\"ottingen
and KIAS on January 2004, August 2004 and March 2005. 
This work was partially supported by JSPS Research Fellowships for
Young Scientists (\#0310363) and the Daiko Foundation (\#9095).



\begin{thebibliography}{0}

\bibitem{ACH}
M.~J.~Ablowitz, S.~Chakravarty and R.~G.~Halburd,
J.\ Math.\ Phys.\  {\bf 44} (2003) 3147.

\bibitem{ACT}
M.~J.~Ablowitz, S.~Chakravarty and L.~A.~Takhtajan,
Commun.\ Math.\ Phys.\  {\bf 158} (1993) 289.

\bibitem{AbCl}
M.~J.~Ablowitz and P.~A.~Clarkson,
{\it Solitons, Nonlinear Evolution Equations and Inverse Scattering},
(Cambridge UP, 1991)
{ [ISBN/0-521-38730-2]}.

\bibitem{AKNS}
M.~J.~Ablowitz, D.~J.~Kaup, A.~C.~Newell and H.~Segur,
Stud.\ Appl.\ Math.\  {\bf 53} (1974) 249.

\bibitem{AGMS}
M.~Aganagic, R.~Gopakumar, S.~Minwalla and A.~Strominger,
JHEP {\bf 0104} (2001) 001
{ [hep-th/0009142]}.

\bibitem{ADHM}
M.~F.~A, N.~J.~Hitchin, V.~G.~Drinfeld and Y.~I.~Manin,
Phys.\ Lett.\ A {\bf 65} (1978) 185.

\bibitem{BBT}
O.~Babelon, D.~Bernard and M.~Talon,
{\it Introduction to classical integrable systems},
(Cambridge UP, 2003)
{ [ISBN/0-521-82267-X]}.

\bibitem{Bak}
D.~Bak,
Phys.\ Lett.\  B {\bf 471} (1999) 149
{ [hep-th/9910135]}.

\bibitem{Bak2}
D.~Bak,
Phys.\ Lett.\ B {\bf 495} (2000) 251
{ [hep-th/0008204]}.

\bibitem{BLP}
  D.~Bak, K.~M.~Lee and J.~H.~Park,
  Phys.\ Rev.\ D {\bf 63} (2001) 125010
  [hep-th/0011099].

\bibitem{BLP2}
D.~s.~Bak, K.~y.~Lee and J.~H.~Park,
Phys.\ Rev.\ D {\bf 66} (2002) 025021
{ [hep-th/0204221]}.

\bibitem{BaDe}
I.~Bakas and D.~A.~Depireux,
Mod.\ Phys.\ Lett.\ A {\bf 6} (1991) 399.

\bibitem{BPST}
A.~A.~Belavin, A.~M.~Polyakov, A.~S.~Schwartz and Y.~S.~Tyupkin,
Phys.\ Lett.\ B {\bf 59} (1975) 85.

\bibitem{Bieling}
S.~Bieling,
J.\ Phys.\ A {\bf 35} (2002) 6281
{ [hep-th/0203269]}.

\bibitem{BCR}
H.~Blas, H.~L.~Carrion and M.~Rojas,
JHEP {\bf 03} (2005) 037
[hep-th/0502051].

\bibitem{Blaszak}
M.~B\l aszak,
{\it Multi-Hamiltonian Theory of Dynamical Systems}
(Springer, 1998) [ISBN/3-540-64251-X].

\bibitem{Cabrera-Carnero}
I.~Cabrera-Carnero,
hep-th/0503147.

\bibitem{CCMo}
I.~Cabrera-Carnero and M.~Moriconi,
Nucl.\ Phys.\ B {\bf 673} (2003) 437
[hep-th/0211193];
hep-th/0303168.

\bibitem{CKN}
S.~Chakravarty, S.~L.~Kent and E.~T.~Newman,
J.\ Math.\ Phys.\  {\bf 36} (1995) 763.


\bibitem{Chu}
C.~S.~Chu,
``Non-commutative geometry from strings,''
hep-th/0502167.

\bibitem{CKT}
C.~S.~Chu, V.~V.~Khoze and G.~Travaglini,
Nucl.\ Phys.\ B {\bf 621} (2002) 101
{ [hep-th/0108007]}.

\bibitem{CFMSS}
  D.~H.~Correa, P.~Forgacs, E.~F.~Moreno, F.~A.~Schaposnik and G.~A.~Silva,
  JHEP {\bf 0407} (2004) 037
  [hep-th/0404015].

\bibitem{CLMS}
D.~H.~Correa, G.~S.~Lozano, E.~F.~Moreno and F.~A.~Schaposnik,
Phys.\ Lett.\ B {\bf 515} (2001) 206
{ [hep-th/0105085]}.

\bibitem{CMS}
D.~H.~Correa, E.~F.~Moreno and F.~A.~Schaposnik,
Phys.\ Lett.\ B {\bf 543} (2002) 235
[hep-th/0207180].

\bibitem{CFZ}
T.~Curtright, D.~Fairlie and C.~K.~Zachos,
Phys.\ Lett.\ B {\bf 405} (1997) 37
{ [hep-th/9704037]};
C.~K.~Zachos, D.~Fairlie and T.~Curtright,
``Matrix membranes and integrability,''
{ hep-th/9709042}.

\bibitem{DJKM}
E.~Date, M.~Jimbo, M.~Kashiwara and T.~Miwa,
``Transformation groups for soliton equations,''
in {\it Nonlinear integrable systems
---classical theory and quantum theory}
(World Sci., 1983) 39 [ISBN/9971950324] and references therein.

\bibitem{Dickey}
L.~A.~Dickey,
{\it Soliton equations and Hamiltonian systems (2nd Ed.)},
Adv.\ Ser.\ Math.\ Phys.\  {\bf 26} (2003) 1 
[TSBN/9812381732].

\bibitem{DiMH}
A.~Dimakis and F.~M\"uller-Hoissen,
Int.\ J.\ Mod.\ Phys.\ B {\bf 14} (2000) 2455
[hep-th/0006005].

\bibitem{DiMH2}
A.~Dimakis and F.~M\"uller-Hoissen,
hep-th/0007015.
Czech.\ J.\ Phys.\ {\bf 51} (2001) 1285.

\bibitem{DiMH3}
A.~Dimakis and F.~M\"uller-Hoissen,
Phys.\ Lett.\ A {\bf 278} (2000) 139
[hep-th/0007074].

\bibitem{DiMH4}
A.~Dimakis and F.~M\"uller-Hoissen,
J.\ Phys.\ A {\bf 34} (2001) 2571
[nlin.si/0008016].

\bibitem{DiMH5}
A.~Dimakis and F.~M\"uller-Hoissen,
Lett.\ Math.\ Phys.\  {\bf 54} (2000) 123
[hep-th/0007160];
J.\ Phys.\ A {\bf 34} (2001) 9163
[nlin.si/0104071].

\bibitem{DiMH6}
A.~Dimakis and F.~M\"uller-Hoissen,
J.\ Phys.\ A {\bf 37} (2004) 4069
[hep-th/0401142];
J.\ Phys.\ A {\bf 37} (2004) 10899 
[hep-th/0406112];
nlin.si/0408023;
J.\ Phys.\ A {\bf 38} (2005) 5453
[nlin.si/0501003].


\bibitem{DHKM}
N.~Dorey, T.~J.~Hollowood, V.~V.~Khoze and M.~P.~Mattis,
Phys.\ Rept.\  {\bf 371} (2002) 231
{ [hep-th/0206063]}.

\bibitem{DoNe}
M.~R.~Douglas and N.~A.~Nekrasov,
Rev.\ Mod.\ Phys.\  {\bf 73} (2002) 977
{ [hep-th/0106048]}.

\bibitem{EINOOST}
  M.~Eto, Y.~Isozumi, M.~Nitta, K.~Ohashi, K.~Ohta and N.~Sakai,
  hep-th/0412024;
  M.~Eto, Y.~Isozumi, M.~Nitta, K.~Ohashi and N.~Sakai,
  hep-th/0412048;
  M.~Eto, Y.~Isozumi, M.~Nitta, K.~Ohashi, K.~Ohta, N.~Sakai and Y.~Tachikawa,
  hep-th/0503033 and references therein.

\bibitem{EGR}
  P.~Etingof, I.~Gelfand and V.~Retakh,
  Math.\ Res.\ Lett.\ {\bf 4} (1997) 413
  [q-alg/9701008].

\bibitem{FJJ}
  O.~Foda, I.~Jack and D.~R.~T.~Jones,
  Phys.\ Lett.\ B {\bf 547} (2002) 79
  [hep-th/0209111].

\bibitem{FSIv}
F.~Franco-Sollova and T.~Ivanova,
J.\ Phys.\ A {\bf 36} (2003) 4207
{ [hep-th/0209153]}.

\bibitem{FIO}
  A.~Fujii, Y.~Imaizumi and N.~Ohta,
  Nucl.\ Phys.\ B {\bf 615} (2001) 61
  [hep-th/0105079].

\bibitem{FINY}
  K.~Furuta, T.~Inami, H.~Nakajima and M.~Yamamoto,
  Phys.\ Lett.\ B {\bf 537} (2002) 165
  [hep-th/0203125];
  JHEP {\bf 0208} (2002) 009
  [hep-th/0207166].

\bibitem{FIY}
K.~Furuta, T.~Inami and M.~Yamamoto,
{ hep-th/0211129}.

\bibitem{Furuuchi}
K.~Furuuchi,
Prog.\ Theor.\ Phys.\ {\bf 103} (2000) 1043
{ [hep-th/9912047]}.

\bibitem{Furuuchi2}
K.~Furuuchi,
Commun.\ Math.\ Phys.\  {\bf 217} (2001) 579
{ [hep-th/0005199]}.

\bibitem{Furuuchi3}
K.~Furuuchi,
{ hep-th/0010006}.

\bibitem{Furuuchi4}
K.~Furuuchi,
JHEP {\bf 0103} (2001) 033
{ [hep-th/0010119]}.

\bibitem{GeDi}
I.~M.~Gelfand and L.~A.~Dikii,
Funct.\ Anal.\ Appl.\ {\bf 10} (1976) 13 (Russian),
259 (English).

\bibitem{GGRW}
I.~Gelfand, S.~Gelfand, V.~Retakh and R.~Wilson,
Adv.\ Math.\ {\bf 193} (2005) 56
[math.QA/0208146].

\bibitem{GeRe}
I.~Gelfand and V.~Retakh,
Funct.\ Anal.\ Appl.\ {\bf 25} (1991) 91; Funct.\ Anal.\ Appl.\
{\bf 2} (1992) 1.

\bibitem{Ghosh}
S.~Ghosh,
Nucl.\ Phys.\ B {\bf 670} (2003) 359
[hep-th/0306045].

\bibitem{GoMa}
  C.~Gomez and J.~J.~Manjarin,
  hep-th/0111169.

\bibitem{GoHa}
  T.~R.~Govindarajan and E.~Harikumar,
  Phys.\ Lett.\ B {\bf 602} (2004) 238
  [hep-th/0406273].

\bibitem{GMPT}
M.~T.~Grisaru, L.~Mazzanti, S.~Penati and L.~Tamassia,
JHEP {\bf 0404} (2004) 057
{ [hep-th/0310214]}.

\bibitem{GrPe}
M.~T.~Grisaru and S.~Penati,
Nucl.\ Phys.\ B {\bf 655} (2003) 250
{ [hep-th/0112246]}.

\bibitem{Groenewold}
H.~J.~Groenewold,
Physica {\bf 12} (1946) 405.

\bibitem{GrNe}
D.~J.~Gross and N.~A.~Nekrasov,
JHEP {\bf 0007} (2000) 034
{ [hep-th/0005204]}.

\bibitem{GrNe2}
D.~J.~Gross and N.~A.~Nekrasov,
JHEP {\bf 0010} (2000) 021
{ [hep-th/0007204]}.

\bibitem{GrNe3}
D.~J.~Gross and N.~A.~Nekrasov,
JHEP {\bf 0103} (2001) 044
{ [hep-th/0010090]}.

\bibitem{Hamanaka}
M.~Hamanaka,
``ADHM/Nahm construction and its duality,''
Soryushiron Kenkyu (Kyoto) {\bf 106} (2002) 1 (Japanese).

\bibitem{Hamanaka2}
M.~Hamanaka,
Phys.\ Rev.\ D {\bf 65} (2002) 085022
{ [hep-th/0109070]}.

\bibitem{Hamanaka3}
M.~Hamanaka, 
``Noncommutative solitons and D-branes,'' 
Ph.\ D thesis (University of Tokyo, 2003) hep-th/0303256.

\bibitem{Hamanaka4}
M.~Hamanaka,
J.\ Math.\ Phys.\  {\bf 46} (2005) 052701 
[hep-th/0311206].

\bibitem{Hamanaka5}
M.~Hamanaka, work in progress.

\bibitem{Hamanaka6}
M.~Hamanaka,
to appear in Phys.\ Lett.\ B [hep-th/0507112].

\bibitem{HIO}
M.~Hamanaka, Y.~Imaizumi and N.~Ohta,
Phys.\ Lett.\ B {\bf 529} (2002) 163
{ [hep-th/0112050]}.

\bibitem{HaTe}
M.~Hamanaka and S.~Terashima,
JHEP {\bf 0103} (2001) 034
{ [hep-th/0010221]}.

\bibitem{HaTo}
M.~Hamanaka and K.~Toda,
Phys.\ Lett.\ A {\bf 316} (2003) 77
{ [hep-th/0211148]}.

\bibitem{HaTo2}
M.~Hamanaka and K.~Toda,
J.\ Phys.\ A {\bf 36} (2003) 11981
{ [hep-th/0301213]}.

\bibitem{HaTo3}
M.~Hamanaka and K.~Toda,
Proc.\ Inst.\ Math.\ NAS Ukraine (2004) 404
[hep-th/0309265].

\bibitem{Hannabuss}
K.~C.~Hannabuss,
Lett.\ Math.\ Phys.\  {\bf 58} (2001) 153
{ [hep-th/0108228]}.

\bibitem{Harvey2}
J.~A.~Harvey,
{ [hep-th/0102076]}.

\bibitem{Harvey3}
J.~A.~Harvey,
{ [hep-th/0105242]}.

\bibitem{HaMo}
J.~A.~Harvey and G.~W.~Moore,
J.\ Math.\ Phys.\  {\bf 42} (2001) 2765
{ [hep-th/0009030]}.

\bibitem{Hashimoto}
K.~Hashimoto,
JHEP {\bf 0012} (2000) 023
{ [hep-th/0010251]}.

\bibitem{Hashimoto2}
  K.~Hashimoto,
  Phys.\ Rev.\ D {\bf 65} (2002) 065014
  [hep-th/0107226].

\bibitem{HaHa}
A.~Hashimoto and K.~Hashimoto,
JHEP {\bf 9911} (1999) 005
{ [hep-th/9909202]}.

\bibitem{HHM}
K.~Hashimoto, H.~Hata and S.~Moriyama,
JHEP {\bf 9912} (1999) 021
{ [hep-th/9910196]}.

\bibitem{HaHi}
K.~Hashimoto and T.~Hirayama,
Nucl.\ Phys.\ B {\bf 587} (2000) 207
{ [hep-th/0002090]}.

\bibitem{HHM2}
K.~Hashimoto, T.~Hirayama and S.~Moriyama,
JHEP {\bf 0011} (2000) 014
{ [hep-th/0010026]}.

\bibitem{HaOo}
K.~Hashimoto and H.~Ooguri,
Phys.\ Rev.\ D {\bf 64} (2001) 106005
 { [hep-th/0105311]}.

\bibitem{Hiraoka}
Y.~Hiraoka,
Phys.\ Lett.\ B {\bf 536} (2002) 147
{ [hep-th/0203047]};
{ hep-th/0205283};
Phys.\ Rev.\ D {\bf 67} (2003) 105025
{ [hep-th/0301176]}.

\bibitem{Ho}
P.~M.~Ho,
{ hep-th/0003012}.

\bibitem{HLW}
Z.~Horv\'ath, O.~Lechtenfeld and M.~Wolf
JHEP {\bf 0212} (2002) 060
{ [hep-th/0211041]}.

\bibitem{IhUh}
M.~Ihl and S.~Uhlmann,
Int.\ J.\ Mod.\ Phys.\ B {\bf 18} (2003) 4889
[hep-th/0211263].

\bibitem{IKS}
T.~Ishikawa, S.~I.~Kuroki and A.~Sako,
JHEP {\bf 0112} (2001) 000
{ [hep-th/0109111]}.

\bibitem{IKS2}
T.~Ishikawa, S.~I.~Kuroki and A.~Sako,
JHEP {\bf 0208} (2002) 028
{ [hep-th/0201196]}.

\bibitem{K.Ito}
K.~Ito,
``Field theory in noncommutative spacetime and superstring theory,''
the Bulletin of the Physical Society of Japan {\bf 59}
(2004) 856 (Japanese).

\bibitem{IvLe}
  T.~A.~Ivanova and O.~Lechtenfeld,
  hep-th/0502117.

\bibitem{ILME}
  T.~A.~Ivanova, O.~Lechtenfeld and H.~Muller-Ebhardt,
  Mod.\ Phys.\ Lett.\ A {\bf 19} (2004) 2419
  [hep-th/0404127].

\bibitem{IvPo}
T.~A.~Ivanova and A.~D.~Popov,
Theor.\ Math.\ Phys.\  {\bf 102} (1995) 280
[Teor.\ Mat.\ Fiz.\  {\bf 102} (1995) 384].

\bibitem{JMW}
  D.~P.~Jatkar, G.~Mandal and S.~R.~Wadia,
  JHEP {\bf 0009} (2000) 018
  [hep-th/0007078].

\bibitem{JiMi}
M.~Jimbo and T.~Miwa,
Publ.\ Res.\ Inst.\ Math.\ Sci.\ Kyoto {\bf 19} (1983) 943.

\bibitem{KKO}
A.~Kapustin, A.~Kuznetsov and D.~Orlov,
Commun.\ Math.\ Phys.\  {\bf 221} (2001) 385
{ [hep-th/0002193]}.

\bibitem{KLY}
K.~Y.~Kim, B.~H.~Lee and H.~S.~Yang,
J.\ Korean Phys.\ Soc.\  {\bf 41} (2002) 290
{ [hep-th/0003093]}.

\bibitem{KLY2}
K.~Y.~Kim, B.~H.~Lee and H.~S.~Yang,
Phys.\ Lett.\ B {\bf 523} (2001) 357
{ [hep-th/0109121]}.

\bibitem{KLY3}
K.~Y.~Kim, B.~H.~Lee and H.~S.~Yang,
Phys.\ Rev.\ D {\bf 66} (2002) 025034
{ [hep-th/0205010]}.

\bibitem{KLY4}
  C.~j.~Kim, K.~M.~Lee and S.~H.~Yi,
  Phys.\ Lett.\ B {\bf 543} (2002) 107
  [hep-th/0204109].

\bibitem{KoSc}
A.~Konechny and A.~Schwarz,
Phys.\ Rept.\  {\bf 360} (2002) 353
{ [hep-th/0012145]}.

\bibitem{KoSc2}
A.~Konechny and A.~Schwarz,
Phys.\ Rept.\  {\bf 360} (2002) 353
{ [hep-th/0107251]}.

\bibitem{KrSh}
  P.~Kraus and M.~Shigemori,
  JHEP {\bf 0206} (2002) 034
  [hep-th/0110035].

\bibitem{Kupershmidt}
B.~Kupershmidt,
{\it KP or mKP}
(AMS, 2000) [ISBN/0821814001].

\bibitem{Lechtenfeld}
  O.~Lechtenfeld,
  Fortsch.\ Phys.\  {\bf 52} (2004) 596
  [hep-th/0401158].

\bibitem{Lechtenfeld2}
O.~Lechtenfeld,
Czech.\ J.\ Phys.\  {\bf 54} (2004) 1351
[hep-th/0409108].

\bibitem{LMPPT}
O.~Lechtenfeld, L.~Mazzanti, S.~Penati, A.~D.~Popov and L.~Tamassia,
Nucl.\ Phys.\ B {\bf 705} (2005) 477
[hep-th/0406065].

\bibitem{LePo}
O.~Lechtenfeld and A.~D.~Popov,
JHEP {\bf 0111} (2001) 040
{ [hep-th/0106213]};
Phys.\ Lett.\ B {\bf 523} (2001) 178
{ [hep-th/0108118]}.

\bibitem{LePo2}
O.~Lechtenfeld and A.~D.~Popov,
JHEP {\bf 0203} (2002) 040
{ [hep-th/0109209]}.

\bibitem{LePo3}
O.~Lechtenfeld and A.~D.~Popov,
JHEP {\bf 0401} (2004) 069
[hep-th/0306263].

\bibitem{LPS}
O.~Lechtenfeld, A.~D.~Popov and B.~Spendig,
Phys.\ Lett.\ B {\bf 507} (2001) 317
{ [hep-th/0012200]};
JHEP {\bf 0106} (2001) 011
{ [hep-th/0103196]}.

\bibitem{LPS2}
  O.~Lechtenfeld, A.~D.~Popov and R.~J.~Szabo,
  JHEP {\bf 0312} (2003) 022
  [hep-th/0310267].


\bibitem{LLY}
  B.~H.~Lee, K.~M.~Lee and H.~S.~Yang,
  Phys.\ Lett.\ B {\bf 498} (2001) 277
  [hep-th/0007140].

\bibitem{LeYa}
B.~H.~Lee and H.~S.~Yang,
Phys.\ Rev.\ D {\bf 66} (2002) 045027
{ [hep-th/0206001]}.

\bibitem{K.Lee}
K.~M.~Lee,
JHEP {\bf 0408} (2004) 054
[hep-th/0405244].

\bibitem{LTY}
K.~M.~Lee, D.~Tong and S.~Yi,
Phys.\ Rev.\ D {\bf 63} (2001) 065017
{ [hep-th/0008092]}.

\bibitem{Legare}
M.~Legare,
{ hep-th/0012077};
J.\ Phys. A {\bf 35} (2002) 5489.

\bibitem{LMS}
  G.~S.~Lozano, E.~F.~Moreno and F.~A.~Schaposnik,
  Phys.\ Lett.\ B {\bf 504} (2001) 117
  [hep-th/0011205].

\bibitem{MaPa}
L.~Martina and O.~K.~Pashaev,
hep-th/0302055.

\bibitem{MMMS}
M.~Marino, R.~Minasian, G.~Moore and A.~Strominger,
JHEP {\bf 0001} (2000) 005
{ [hep-th/9911206]}.

\bibitem{MaSp}
  L.~J.~Mason and G.~A.~J.~Sparling,
  Phys.\ Lett.\ A {\bf 137} (1989) 29.

\bibitem{MaWo}
L.~J.~Mason and N.~M.~Woodhouse,
{\it Integrability, Self-Duality, and Twistor Theory}
(Oxford UP, 1996)
{ [ISBN/0-19-853498-1]}.

\bibitem{Matsuo}
Y.~Matsuo,
Phys.\ Lett.\ B {\bf 499} (2001) 223
{ [hep-th/0009002]}.

\bibitem{MPT}
  M.~Mihailescu, I.~Y.~Park and T.~A.~Tran,
  Phys.\ Rev.\ D {\bf 64} (2001) 046006
  [hep-th/0011079].

\bibitem{DJM}
T.~Miwa, M.~Jimbo and  E.~Date, (translated by M.~Reid),
``Solitons : differential equations,
symmetries and infinite dimensional algebras,''
(Cambridge UP, 2000)
{ [ISBN/0521561612]}.

\bibitem{Moriyama}
S.~Moriyama,
Phys.\ Lett.\ B {\bf 485} (2000) 278
{ [hep-th/0003231]}.

\bibitem{Moriyama2}
S.~Moriyama,
JHEP {\bf 0008} (2000) 014
{ [hep-th/0006056]}.

\bibitem{Moyal}
J.~E.~Moyal,
Proc.\ Cambridge Phil.\ Soc.\  {\bf 45} (1949) 99;

\bibitem{AdMu}
  J.~Murugan and R.~Adams,
  JHEP {\bf 0212} (2002) 073
  [hep-th/0211171].

\bibitem{Nakajima2}
H.~Nakajima,
``Resolutions of moduli spaces of ideal instantons on ${\bf R}^4$,''
{\it Topology, Geometry and Field Theory}
(World Sci., 1994) 129
{ [ISBN/981-02-1817-6]}.

\bibitem{Nakajima3}
H.~Nakajima,
{\it Lectures on Hilbert Schemes of Points on Surfaces}
(AMS, 1999) { [ISBN/0-8218-1956-9]}.

\bibitem{Nekrasov}
N.~A.~Nekrasov,
Commun.\ Math.\ Phys.\  {\bf 241} (2003) 143
{ [hep-th/0010017]}.

\bibitem{Nekrasov2}
N.~A.~Nekrasov,
{ hep-th/0011095}.

\bibitem{Nekrasov3}
N.~A.~Nekrasov,
{ hep-th/0203109}.

\bibitem{NeSc}
N.~Nekrasov and A.~Schwarz,
Commun.\ Math.\ Phys.\ {\bf 198} (1998) 689
{ [hep-th/9802068]}.

\bibitem{NiRa}
H.~Nishino and S.~Rajpoot,
Phys.\ Lett.\ B {\bf 572} (2003) 91
[hep-th/0306290].

\bibitem{Ohta_k}
K.~Ohta,
Phys.\ Rev.\ D {\bf 64} (2001) 046003
{ [hep-th/0101082]}.

\bibitem{OSTT}
Y.~Ohta, J.~Satsuma, D.~Takahashi and T.~Tokihiro,
Prog.\ Theor.\ Phys.\ Suppl.\  {\bf 94} (1988) 210.

\bibitem{OoVa}
H.~Ooguri and C.~Vafa,
Mod.\ Phys.\ Lett.\ A {\bf 5} (1990) 1389;
Nucl.\ Phys.\ B {\bf 361} (1991) 469;
Nucl.\ Phys.\ B {\bf 367} (1991) 83.

\bibitem{IKOS}
  H.~Otsu, T.~Sato, H.~Ikemori and S.~Kitakado,
  JHEP {\bf 0307} (2003) 054
  [hep-th/0303090];
  JHEP {\bf 0406} (2004) 006
  [hep-th/0404140];
  hep-th/0503118.

\bibitem{Paniak}
L.~D.~Paniak,
{ hep-th/0105185}.

\bibitem{Parvizi}
S.~Parvizi,
  Mod.\ Phys.\ Lett.\ A {\bf 17} (2002) 341
  [hep-th/0202025].

\bibitem{Penrose}
R.~Penrose,
Gen.\ Rel.\ Grav.\  {\bf 7} (1976) 31.

\bibitem{Polychronakos}
A.~P.~Polychronakos,
Phys.\ Lett.\ B {\bf 495} (2000) 407
{ [hep-th/0007043]}.

\bibitem{PSW}
A.~D.~Popov, A.~G.~Sergeev and M.~Wolf,
J.\ Math.\ Phys.\  {\bf 44} (2003) 4527
[hep-th/0304263].

\bibitem{PoSz}
  A.~D.~Popov and R.~J.~Szabo,
  hep-th/0504025.

\bibitem{Sakakibara}
M.~Sakakibara,
J.\ Phys.\ A {\bf 37} (2004) L599
[nlin.si/0408002].

\bibitem{Sako}
A.~Sako,
JHEP {\bf 0304} (2003) 023
{ [hep-th/0209139]}.

\bibitem{SaSu}
  A.~Sako and T.~Suzuki,
  hep-th/0503214.

\bibitem{SaSa}
M.~Sato and Y.~Sato,
``Soliton equations as dynamical systems on infinite
dimensional Grassmann manifold,''
in {\it Nonlinear Partial Differential Equations in Applied Sciences}
(North-Holland, 1982) 259.

\bibitem{Schaposnik}
  F.~A.~Schaposnik,
  Braz.\ J.\ Phys.\  {\bf 34} (2004) 1349
  [hep-th/0310202];
  hep-th/0408132.

\bibitem{Schwarz}
A.~Schwarz,
Commun.\ Math.\ Phys.\  {\bf 221} (2001) 433
{ [hep-th/0102182]}.

\bibitem{Schwarz2}
  A.~Schwarz,
  hep-th/0111174.

\bibitem{SeWi}
N.~Seiberg and E.~Witten,
JHEP {\bf 9909} (1999) 032
[hep-th/9908142].

\bibitem{Sen}
  A.~Sen,
  hep-th/0410103.

\bibitem{SWW}
K.~Shigechi, M.~Wadati and N.~Wang,
Nucl.\ Phys.\ B {\bf 706} (2005) 518
[hep-th/0404249].

\bibitem{Strachan}
  I.~A.~B.~Strachan,
  J.\ Geom.\ Phys.\ {\bf 21} (1997) 255
  [hep-th/9604142].

\bibitem{Szabo}
R.~J.~Szabo,
Phys.\ Rept.\  {\bf 378} (2003) 207
{ [hep-th/0109162]}.

\bibitem{Takasaki}
K.~Takasaki,
J.\ Geom.\ Phys.\ {\bf 37} (2001) 291
{ [hep-th/0005194]}.

\bibitem{Terashima}
S.~Terashima,
Phys.\ Lett.\ B {\bf 477} (2000) 292
{ [hep-th/9911245]}.

\bibitem{Tian}
Y.~Tian,
hep-th/0307264.

\bibitem{Tian2}
  Y.~Tian,
  Mod.\ Phys.\ Lett.\ A {\bf 19} (2004) 1315
  [hep-th/0404128].

\bibitem{TiZh}
Y.~Tian and C.~J.~Zhu,
Commun.\ Theor.\ Phys.\  {\bf 38} (2002) 691
{ [hep-th/0205110]}.
Phys.\ Rev.\ D {\bf 67} (2003) 045016
{ [hep-th/0210163]}.

\bibitem{STZ}
Y.~Tian, C.~J.~Zhu and X.~C.~Song,
Mod.\ Phys.\ Lett.\ A {\bf 18} (2003) 1691
{ [hep-th/0211225]}.

\bibitem{Toda}
K.~Toda,
Proceedings of workshop on
Integrable Theories, Solitons and Duality, Sao Paulo,
Brazil, 1-6 July 2002 { [JHEP PRHEP-unesp2002/038]}.

\bibitem{Valtancoli}
  P.~Valtancoli,
  Int.\ J.\ Mod.\ Phys.\ A {\bf 18} (2003) 1125
  [hep-th/0209118].

\bibitem{WaWa}
N.~Wang and M.~Wadati,
J.\ Phys.\ Soc.\ Jap.\ {\bf 72} (2003) 1366;
J.\ Phys.\ Soc.\ Jap.\ {\bf 72} (2003) 1881.

\bibitem{WaWa2}
N.~Wang and M.~Wadati,
J.\ Phys.\ Soc.\ Jap.\ {\bf 72} (2003) 3055.

\bibitem{WaWa3}
N.~Wang and M.~Wadati,
J.\ Phys.\ Soc.\ Jap.\  {\bf 73} (2004) 1689.

\bibitem{Ward}
R.~S.~Ward,
Phil.\ Trans.\ Roy.\ Soc.\ Lond.\ A {\bf 315} (1985) 451;
Lect.\ Notes Phys.\ {\bf 280} (Springer, 1986) 106;
``Integrable systems in twistor theory,''
in {\it Twistors in Mathematics and Physics}
(Cambridge UP, 1990) 246.

\bibitem{WaWe}
R.~S.~Ward and R.~O.~Wells, {\it Twistor Geometry and Field
Theory} (Cambridge UP, 1990) [ISBN/0-521-42268-X].

\bibitem{Watamura}
S.~Watamura,
``Gauge theories on noncommutative spaces,''
Proceedings of workshop on Mathematical Physics 2002,
Tokyo, Japan, 21-23 September 2002 (Japanese).

\bibitem{Wimmer}
  R.~Wimmer,
  JHEP {\bf 0505} (2005) 022
  [hep-th/0502158].

\bibitem{Witten}
  E.~Witten,
  Phys.\ Rev.\ Lett.\  {\bf 38} (1977) 121.

\bibitem{Witten2}
  E.~Witten,
  JHEP {\bf 0204} (2002) 012
  [hep-th/0012054].

\bibitem{Wolf}
M.~Wolf,
JHEP {\bf 0206} (2002) 055
{ [hep-th/0204185]}.

\bibitem{Zuevsky}
A.~Zuevsky,
J.\ Phys.\ A {\bf 37} (2004) 537.

\end{thebibliography}
\end{document}